# Benchmarking the Open Science Data Federation services to develop XRootD best practices

*Fabio* Andrijauskas*, *Igor* Sfiligoi, and *Frank* Würthwein
University of California San Diego, San Diego Supercomputer Center, 9836 Hopkins Dr, La Jolla, CA 92093 USA

**Abstract.** Research has become dependent on processing power and storage, one crucial aspect being data sharing. The Open Science Data Federation (OSDF) project aims to create a scientific global data distribution network based on the Pelican Platform. OSDF relies on the XRootD and Pelican projects. Nevertheless, OSDF must understand the XRootD limits under various configuration options, including transfer rate limits, proper buffer configuration, and storage type effect. We have thus executed a set of benchmarks to create a set of recommendations to share with the XRootD and Pelican teams. This work describes the tests and results performed using National Research Platform (NRP) hosts. The tests cover various file sizes and parallel streams and use clients from various distances from the server host. We also used several standalone clients (*wget, curl*, pelican) and the native HTCondor file transfer mechanisms. Applying the methodology creates a possibility to track how XRootD and the Pelican layer perform in different scenarios.

## 1 Introduction

Research in today's digital landscape heavily relies on advanced processing capabilities and vast storage solutions. A particularly critical aspect is facilitating data sharing among scientists and maintaining all the necessary accounting standards and proper units for computational resources [1]. The Open Science Data Federation (OSDF) [2] project, a crucial initiative in this landscape, is designed to establish a comprehensive global data distribution network that enhances collaboration and accessibility in the scientific community.

While OSDF does not engage in software development, it leverages existing frameworks such as XrootD [3] and Pelican Platform (https://pelicanplatform.org) to achieve its objectives [4]. These frameworks, particularly XRootD, play a fundamental role in OSDF's mission. Thus, it is essential to understand XRootD's limitations under different configurations. This includes investigating critical factors such as transfer rate

* Corresponding author: fandrijauskas@ucsd.edu

caps, optimal buffer configurations, and the impact of various storage types on performance.

We conducted a comprehensive set of benchmarks to address these objectives and formulate targeted recommendations to support the XRootD and Pelican Platform development teams. This document details the methodologies used in our tests and the results obtained while utilizing hosts from the National Research Platform (NRP). Our benchmarking covered a diverse range of file sizes and assessed multiple parallel streams, considering client access from various geographical distances relative to the server. Additionally, we employed several standalone client tools, including wget, *curl*, and Pelican, alongside the native file transfer mechanisms provided by HTCondor. Through this rigorous testing, we aim to provide valuable insights that will enhance the efficiency and reliability of data sharing via the OSDF initiative. This paper aims to create and apply a methodology to test XRootD in a production environment and track XRootD performance along with new versions and features.

## 2 Methodology

This methodology aims to check basic XRootD features, such as retrieving a file from a server, using XRootD as a cache, and other specific features, such as overhead during the authentication process. Each item has an impact (1 to 5) and complexity (1 to 5) to be completed. The impact is related to improving the service availability, security, new features, user experience, etc. The complexity is related to the required time to create the test case and the time to run the tests. Tables 1 to 10 show the proposed XRootD tests' complexity, impact, and description. The methodology was designed to verify the data integrity and generate a comprehensive dataset utilizing OSDF hosts. This process involved a series of systematic tests labeled as tests 1, 2, 3, and 4. Tests 6 to 10 are described as a reference for a complete XRootD benchmark and used to test other scenarios, plugins, and versions in future works.

Each test assessed different aspects of the OSDF hosts, ensuring that the generated dataset met the standards for accuracy and reliability. Through careful execution and analysis of these tests, we gathered valuable insights and produced a robust dataset suitable for further research and application.

**Table 1.** File transfer from an origin to a client using six file sizes and 1, 8, 32, 64, 128 to N streams.

| 1 | File transfer from an origin to a client. \| **Impact:** 5 \| **Complexity:** 3 |
|---|---|
| **Description.** Test the transfer rate using six file sizes (1KB, 1MB, 100MB, 1GB, 10GB, 100GB) using 1, 8, 32, 64, 128 to N streams (where N is the number of parallel transfers) in the LAN and the WAN, using at least three significantly different RTT values. Check throughput for various RTTs and some async settings. This will inform us if we should make this more configurable, either through opaque parameters or automatically, based on detected RTT also, check the transfer rate using different clients (wget, curl, pelican) and HTCondor jobs. This set of tests should be able to create a transfer rate base. | |

**Table 2.** High load on CPU during 24 hours.

| 2 | How XRootD behaves with a High load on CPU and IO for 24 hours. \| **Impact:** 3 \| **Complexity:** 3 |
|---|---|

| **Description** Check the number of errors or problems on the logs and on the requests with file requests using six file sizes (1KB, 1MB, 100MB, 1GB, 10GB, 100GB), document how the main storage is mounted, check IO load, and other software configurations. |
|---|

**Table 3.** Check the best resource configuration between K8S and XRootD

| **3** | Check the best resource configuration between K8S Pod, host resources, and XRootD parameters. \| **Impact:** 2 \| **Complexity:** 3 |
|---|---|
| **Description:** Checking the balance between the host resources and the POD resources using different kinds of tests. | |

**Table 4.** Check the transfer rate difference between the origin access and the closest cache.

| **4** | Check the transfer rate difference between an origin access and the closest cache. \| **Complexity:** 3 \| **Impact:** 3 |
|---|---|
| **Description**. Check the transfer rate difference between origin access and the closest cache, using six file sizes (1KB, 1MB, 100MB, 1GB, 10GB, 100GB) and test the evict function on the cache. | |

**Table 5.** Check the transfer rate related to the storage type.

| **5** | Check the transfer rate related to the storage type. \| **Complexity:** 3 \| **Impact:** 5 |
|---|---|
| **Description** Test the transfer rate using SSD and HDD (RAID and accessing the driver directly) using six file sizes (1KB, 1MB, 100MB, 1GB, 10GB, 100GB) using 1, 2, 4, and 8 streams. | |

**Table 6.** Check the impact of authenticated and unauthenticated access.

| **6** | Check the transfer rate between an authenticated and unauthenticated. \| **Complexity:** 3 \| **Impact:** 5 |
|---|---|
| **Description**. Check the transfer rate between an authenticated and unauthenticated, using tokens, CVMFS (auth or not), or certificate and using six file sizes (1KB, 1MB, 100MB, 1GB, 10GB, 100GB) using 1, 2, 4, and 8 streams. | |

**Table 7.** Check the overhead of token usage for file transfers.

| **7** | Check the overhead of tokens. \| **Complexity:** 3 \| **Impact:** 5 |
|---|---|
| **Description**. Check the overhead of tokens, generate X unique tokens to avoid cache and see how quickly XRootD can authorize them. | |

**Table 8.** Check the transfer rate using HTTP Third party copy.

| **8** | Check the transfer rate using HTTP Third party copy. \| **Complexity:** 3. \| **Impact:** 5 |
|---|---|
| Check the transfer rate using HTTP Third party copy using tokens or certificates and using six file sizes (1KB, 1MB, 100MB, 1GB, 10GB, 100GB) using 1, 2, 4, and 8 streams. | |

**Table 9.** Check the performance of EL7 vs EL9.

| **9** | Check the performance of EL7 vs EL9 as K8S OS POD. \| **Complexity:** 3 \| **Impact:** 4 |
|---|---|
| **Description**. Test the transfer rate using six file sizes (1KB, 1MB, 100MB, 1GB, 10GB, 100GB) using 1, 2, 4, and 8 streams and EL7 vs EL9 as K8S OS POD. | |

**Table 10.** Check how the redirector performs and other elements.

| **10** | Check how the redirector performs, as well as the architecture, and improve the monitoring. \| **Complexity: 3 \| Impact:** 4 |
|---|---|

**Description.** Test the transfer rate using six file sizes (1KB, 1MB, 100MB, 1GB, 10GB, 100GB) using 1, 2, 4, and 8 streams and EL7 vs EL9 as K8S OS POD forcing the redirector be used in each transfer.

## 3 Results and discussion

The methodology was applied to test 2 hosts used by the OSDF, one cache and one origin. The request to this cache and origin originated from 3 other hosts in San Diego – CA, Chicago – IL, and Jacksonville – FL. Table 11 shows the distance and RTT between the three hosts. Table 1 shows the RTT from the Origin and cache for the test points; the numbers are *min/avg/max/sdev* (standard deviation) and the distance in km. The hardware chosen is an attempt to create a homogenous scenario. All the results require interpretation due to the hardware difference.

**Table 11.** RTT from the origin and cache.

| | Origin | Cache |
|---|---|---|
| San Diego | 0.088/0.109/0.186/0.038 ms<br>0 km | 0.066/0.178/0.339/0.092 ms<br>0 km |
| Chicago | 47.331/47.350/47.391/0.023 ms<br>2,784.10 km | 47.337/47.353/47.394/0.021 ms<br>2,784.10 km |
| Jacksonville | 51.324/56.352/57.381/0.023 ms<br>3,359.86 km | 51.325/56.354/57.383/0.023 ms<br>3,359.86 km |

Close to 11,000 minutes were dedicated to transferring and executing tests. Additionally, numerous commits were made to the Tiger repository during this process.

### 3.1 Test cases 1 and 4: File transfer from an origin/cache to a client – Regular Posix files

The image in Figure 1 displays the outcomes of accessing the target origin and cache from the San Diego, California, test point for 15 minutes for each number of parallel threads and file sizes; at the end of the test, the transfer rate is calculated using the time used and the file download size.

It showcases various file sizes and parallel requests while utilizing regular HTTP access. Transfer using the cache is two times faster than the origin, using file sizes of 100GB, 10GB, 1GB, and 100MB; the origin is faster for files of 1KB. In this case, the physical proximity decreased the discrepancy between the cache and origin transfer rate. The tests were conducted using all the machines described in Table 11. Figure 1 illustrates the findings from accessing the target origin and cache from the Jacksonville, Florida, test point. The results highlight the impact of utilizing various file sizes and parallel requests through regular HTTP access. This test shows a better performance with fewer threads.

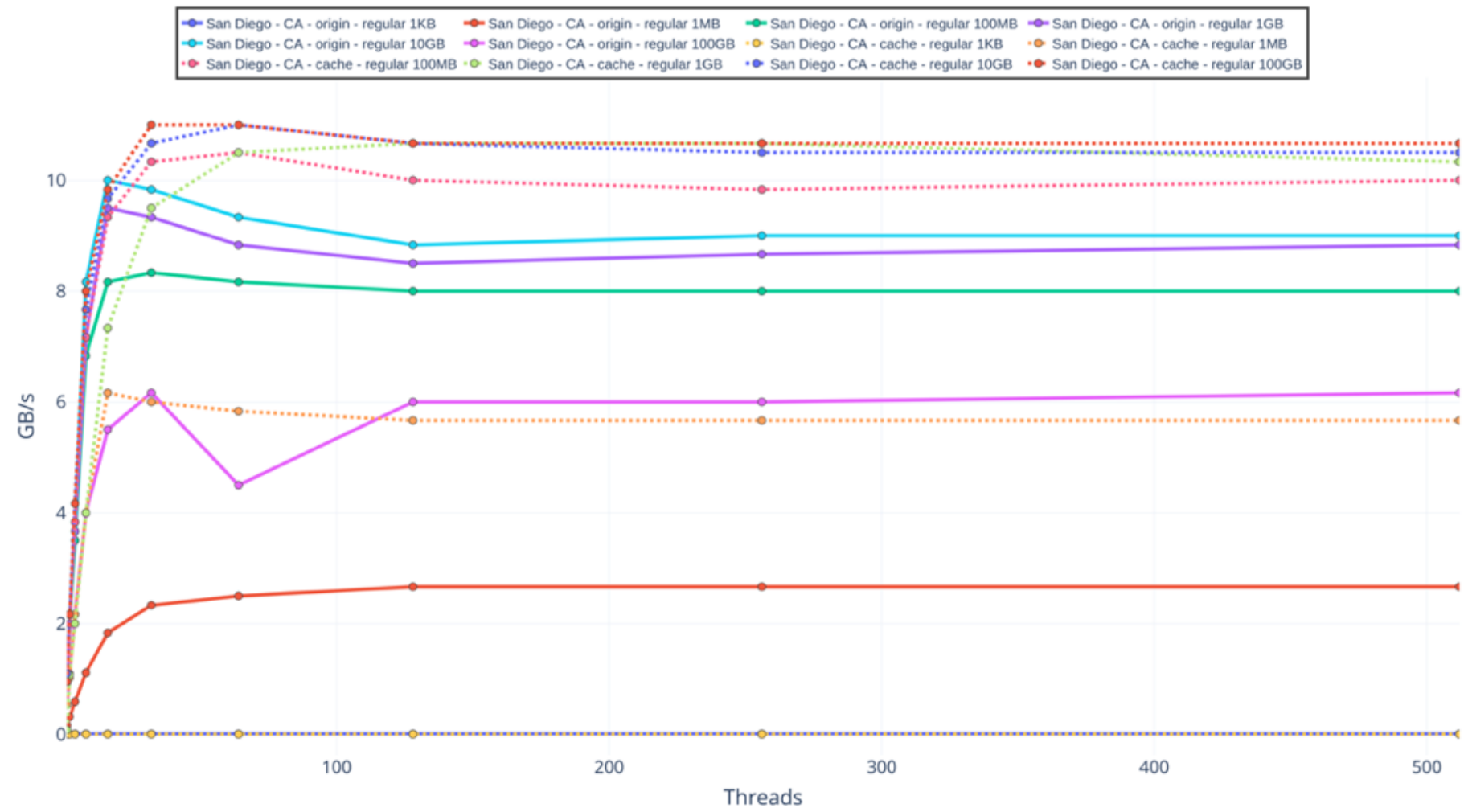

**Fig 1.** Testing the cache and the origin using the San Diego test point using regular HTTP access.

Figure 2 illustrates the findings obtained from accessing the target origin and cache from the Jacksonville, Florida, test point for 15 minutes for each number of parallel threads and file sizes; at the end of the test, the transfer rate is calculated using the time used and the file download size. The results highlight the impact of utilizing various file sizes and parallel requests through regular HTTP access. This test shows a better performance with fewer threads.

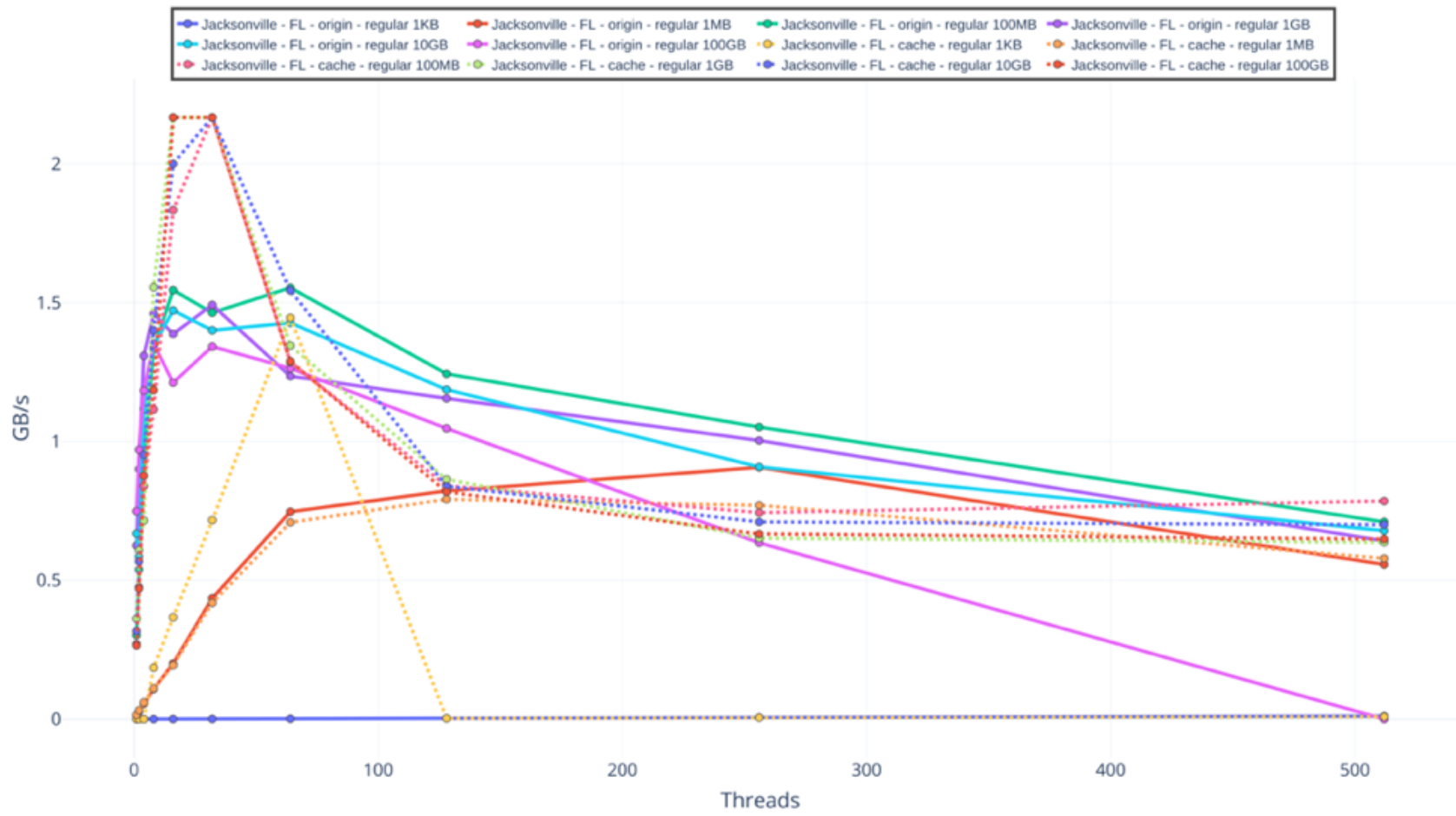

**Fig 2.** Testing the cache and the origin using Jacksonville - FL test point using regular HTTP access.

### 3.2 Test case 2: How XRootD behaves with a High load on CPU and IO for 24 hours.

We encountered some infrastructure issues during testing and identified a bug during high-load testing in Pelican. These challenges underscore the need for further investigation and potential improvements in the Pelican origin and cache deployment. Below are the specific issues recorded for Pelican 7.10 and 7.9. This result shows the importance of a load test for each Pelican/XRootD release.

### 3.3 How OSDF/Pelican clients perform

The performance of the Pelican client and the regular wget was measured and compared across various file sizes, taking 15 minutes for each file size; at the end of the test, the transfer rate was calculated using the time used and the file download size. The specific outcomes of this comparison can be found in Figure 3. There is no statistical difference in the file size used for the transfer.

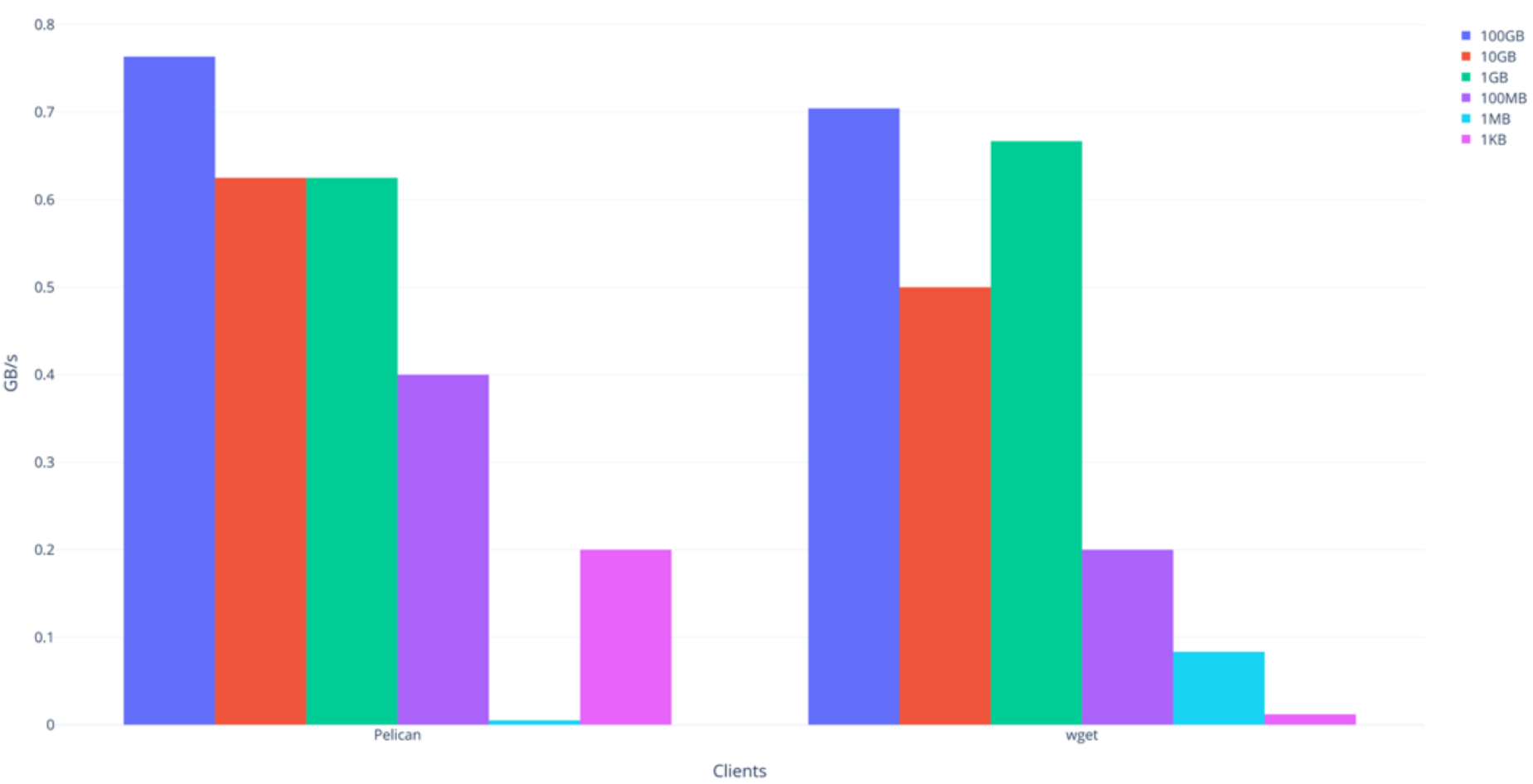

**Fig 3.** Performance testing between Wget and Pelican client.

### 3.4 Test case 3: Check the best resource configuration between K8S Pod, host resources, and XRootD parameters.

The graph in Figures 4 and 5 illustrates the relationship between the number of cores utilized in K8S and the corresponding transfer rate performance. The tests were performed for 15 minutes for each number of parallel threads and the number of cores; at the end of the test, the transfer rate was calculated using the time used and the file download size. More cores represent a speedup for larger files (100GB, 10GB, and 1GB). However, reaching the same transfer rate is possible using fewer (55 and 44 cores); using 20 or less degrades performance. For small files (1KB, 1MB, and 100MB), more cores do not correlate with a faster transfer rate.

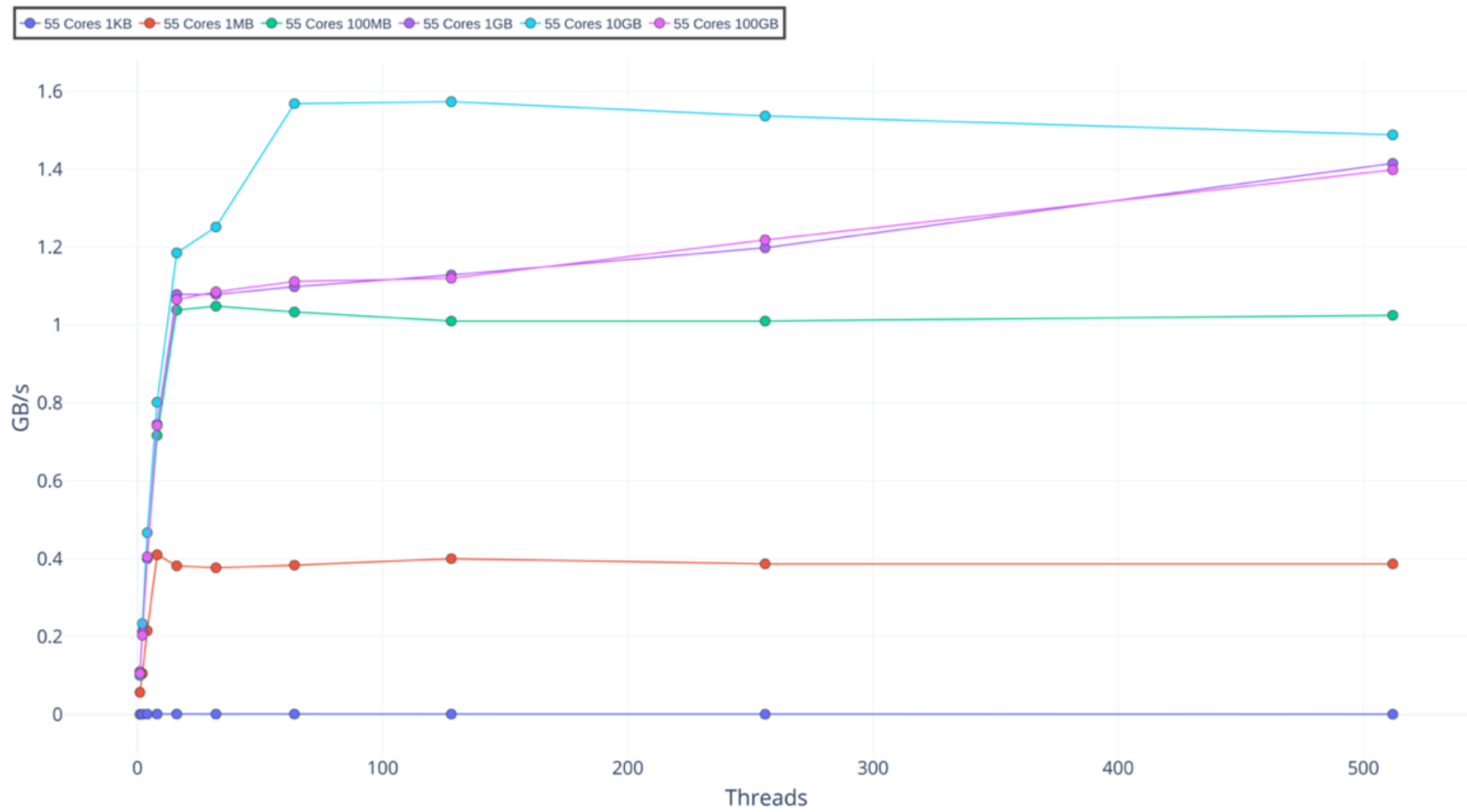

**Fig 4.** Tests were conducted using different file sizes (1KB, 1MB, 100MB, 1GB, 10GB, and 100GB) with 55 cores in a K8S POD with 64 cores in total.

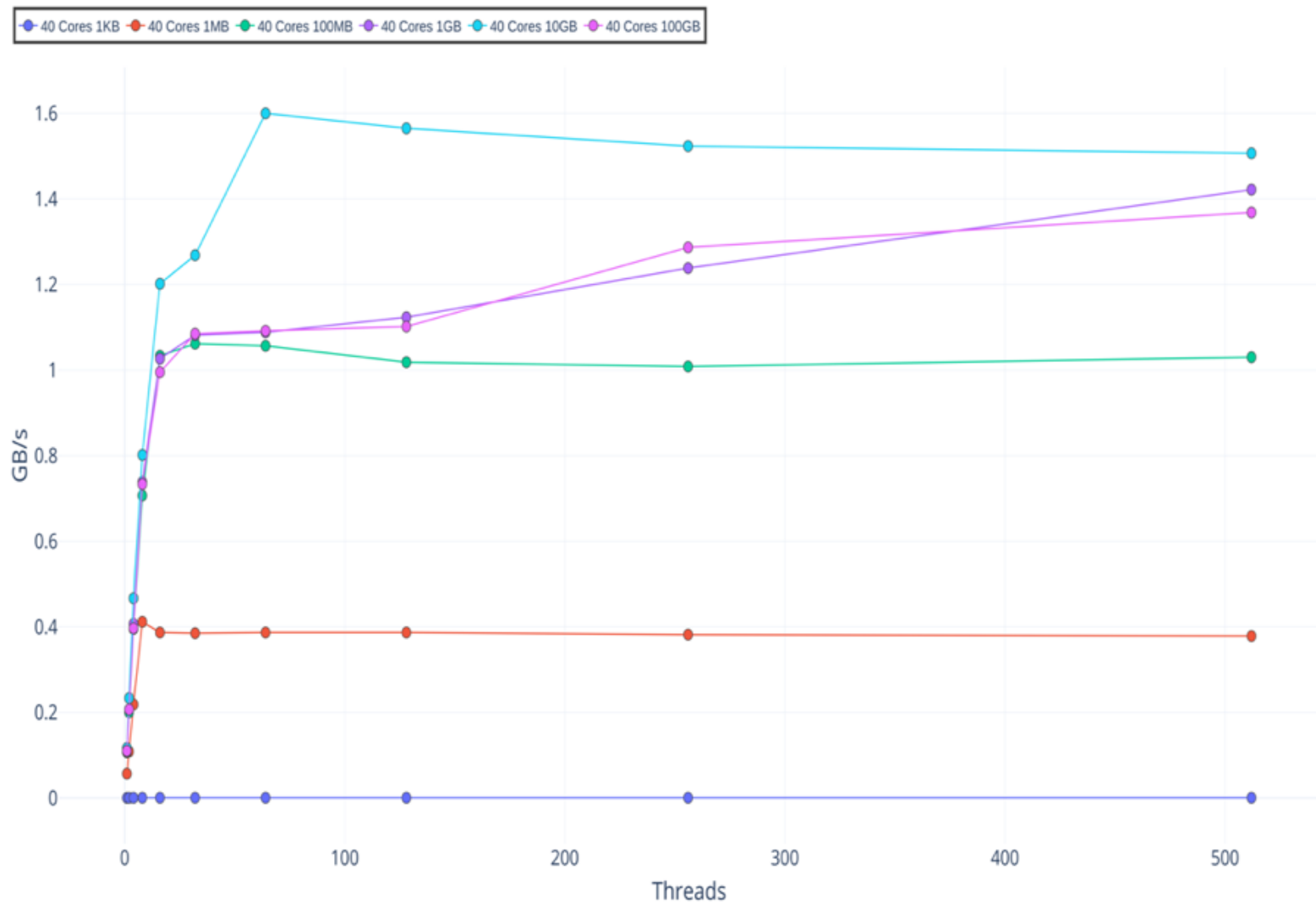

**Fig 5.** Tests were conducted using different file sizes (1KB, 1MB, 100MB, 1GB, 10GB, and 100GB) with 40 cores in a K8S POD with 64 cores in total.

## Conclusion

All software requires benchmarking to some extent. The XRootD and Pelican platforms are used globally, making it essential to develop a standardized testing methodology. When using the Pelican client, checking if the file is present at the origin in case a 404 error occurs is essential. The number of CPUs in the K8S setup should be determined based on the sizes of the requested files. Caches significantly improve the transfer process, reducing transfer times by at least threefold. Conversely, regular access to OSDF is more efficient with larger files (1GB to 100GB) than with small files (1MB to 100MB). Comparatively, no discernible statistical difference exists between using the Pelican client and a regular HTTP client. Further testing of the token and authenticated access is essential to isolate the source of the 404 error and node and container reboot issues. Performance-wise, there are no significant statistical differences when using the Pelican client versus a regular HTTP client in both authenticated and non-authenticated scenarios. The presence of log rotation in the OSDF docker images is a critical factor. OSDF is optimized to transfer larger files, so configuration tuning to small files is required. To provide access to small files, creating " micro-caches " that are optimized for small files and multiple requests is possible.

## Acknowledgment

This work was partially supported by the NSF grants OAC-2112167, OAC-2030508, OAC-1841530, OAC-1836650, the CC* program, and in-kind contributions by many institutions, including ESnet, Internet2, and the Great Plains Network.